\documentstyle[aps,epsfig,floats]{revtex}

\begin{document}

\wideabs{
\title{The anisotropic Heisenberg model in a transverse magnetic field}
\author{D.V.Dmitriev$^{a,b}$, V.Ya.Krivnov$^{a,b}$, A.A.Ovchinnikov$^{a,b}$ and
A.Langari$^{b,c}$}
\address{$^a$ Joint Institute of Chemical Physics of RAS,
Kosygin str.4, 117977, Moscow, Russia \\
$^b$ Max-Planck-Institut fur Physik Komplexer Systeme, Nothnitzer
Str. 38, 01187 Dresden, Germany\\
$^c$ Institute for Advanced Studies in Basic Sciences, Zanjan
45195-159, Iran}
\maketitle
\begin{abstract}
One dimensional spin-$1/2$ $XXZ$ model in a transverse magnetic
field is studied. It is shown that the field induces the gap in
the spectrum of the model with easy-plain anisotropy. Using
conformal invariance the field dependence of the gap at small
fields is found. The ground state phase diagram is obtained. It
contains four phases with different types of the long range order
(LRO) and a disordered one. These phases are separated by critical
lines, where the gap and the long range order vanish. Using
scaling estimations and a mean-field approach as well as numerical
calculations in the vicinity of all critical lines we found the
critical exponents of the gap and the LRO. It is shown that
transition line between the ordered and disordered phases belongs
to the universality class of the transverse Ising model.
 75.10.Jm - Quantized spin models
\end{abstract}
} 

The effect of a magnetic field on an antiferromagnetic chain has
been attracting much interest from theoretical and experimental
points of view. In particular, the strong dependence of the
properties of quasi-one-dimensional anisotropic antiferromagnets
on the field orientation was experimentally observed \cite{YbAs}.
So, it is interesting to study the dependence of properties of the
one-dimensional antiferromagnet on the direction of the applied
field. The simplest model of the one-dimensional anisotropic
antiferromagnet is the spin-$1/2$ $XXZ$ model. This model in an
uniform longitudinal magnetic field (along the $Z$ axis) was
studied in a great details \cite {Yang}. Since the longitudinal
field commutes with the $XXZ$ Hamiltonian the model can be exactly
solved by the Bethe ansatz. This is not the case when the
symmetry-breaking transverse magnetic field is applied and the
exact integrability is lost. Because of a mathematical complexity
of this model it has not been studied so much. From this point of
view it is of a particular interest to study the ground state
properties of the 1D $XXZ$ model in the transverse magnetic field.
The Hamiltonian of this model reads
\begin{equation}
H=\sum_{n=1}^{N}(S_{n}^{x}S_{n+1}^{x}+S_{n}^{y}S_{n+1}^{y}+\Delta
S_{n}^{z}S_{n+1}^{z})+h\sum_{n=1}^{N}S_{n}^{x}  \label{H}
\end{equation}
with periodic boundary conditions and even $N$.

The spectrum of the $XXZ$ model for $-1<\Delta \leq 1$ is gapless.
In the longitudinal field the spectrum remains gapless if the
field does not exceed a saturated value $1+\Delta $. On the
contrary, when the transverse magnetic field is applied a gap in
the excitation spectrum seems to open up. It is supposed
\cite{kufo} that this effect can explain the peculiarity of low
temperature specific heat in ${\rm Yb}_{4}{\rm As}_{3}$
\cite{YbAs}. The magnetic properties of this compound is described
by $XXZ$ Hamiltonian with $\Delta \approx 0.98$ and it was shown
\cite{kufo} that the magnetic field in easy plain induces a gap in
the spectrum resulting in a dramatic decrease of the linear term
in the specific heat.

First of all, what do we know about the model (\ref{H})?

The first part of the Hamiltonian is well-known $XXZ$ model with
the exact solution given by Bethe ansatz. In the Ising-like region
$\Delta >1$ the ground state of $XXZ$ model has a Neel long-range
order (LRO) along $Z$ axis and there is a gap in excitation
spectrum. In the region $-1<\Delta \leq 1$ the system is in the
so-called spin-liquid phase with a power-law decay of correlations
and a linear spectrum. And, finally, for $\Delta <-1$ classical
ferromagnetic state is a ground state of $XXZ$ model with a gap
over ferromagnetic state.

In the transverse magnetic field the total $S^{z}$ is not a good
quantum number and the model is essentially complicated, because
the transverse field breaks rotational symmetry in $X-Y$ plane and
destroys the integrability of the $XXZ$ model, except some special
points. In particular, the exact diagonalization study of this
model is difficult for finite systems because of non-monotonic
behavior of energy levels.

The first special case of the model (\ref{H}) is the limit $\Delta
\rightarrow \pm \infty $. In this case the model (\ref{H}) reduces
to the 1D Ising model in a transverse field (ITF), which can be
solved exactly by transformation to the system of non-interacting
fermions. In both limits the system has a phase transition point
$h_{c}=|\Delta |/2$, where the gap is closed and the LRO in $Z$
direction vanishes.

It is suggested \cite{J>0} that the phase transition of the ITF
type takes place for any $\Delta >0$ at some critical value
$h=h_{c}(\Delta )$. One can expect also that such a transition
exists for any $\Delta $ and the transition line connects two
limiting points $h_{c}=|\Delta |/2$, $\Delta \rightarrow \pm
\infty $.

Similar to these limiting cases, for any $|\Delta |>1$ and
$h<h_{c}(\Delta )$ the system has the LRO in $Z$ direction (Neel
for $\Delta >1$ and ferromagnetic for $\Delta <-1$). But for
$|\Delta |<1$ and $h<h_{c}(\Delta )$ the ground state changes and
instead of the LRO in $Z$ direction a staggered magnetization
along $Y$ axis appears at $h<h_{c}(\Delta )$.

This assumption is confirmed on the `classical' line $h_{{\rm
cl}}=\sqrt{2(1+\Delta )}$ ($h_{{\rm cl}}<h_{c}(\Delta )$), where
the quantum fluctuations of $XXZ$ model are compensated by the
transverse field and the exact ground state of (\ref{H}) at
$h=h_{{\rm cl}}$ is a classical one \cite {classical}. The excited
states on the classical line are generally unknown, though it is
assumed that the spectrum is gapped.

The second case, where the model (\ref{H}) remains integrable, is
the isotropic AF case $\Delta =1$. In this case the direction of
the magnetic field is not important and the ground state of the
system remains spin-liquid one up to the point $h=2$, where the
phase transition of the Pokrovsky-Talapov type takes place and the
ground state becomes completely ordered ferromagnetic state.

And the last special case is $\Delta =-1$. In this case the model
(\ref{H}) reduces to the isotropic ferromagnetic model in a
staggered magnetic field. This model is non-integrable, but as was
shown \cite{Alkaraz}, the system remains gapless up to some
critical value $h=h_{0}$, where the phase transition of the
Kosterlitz-Thouless type takes place.

Summarizing all above, we expect that the phase diagram of the
model (\ref{H}) (on ($\Delta $, $h$) plane) has a form as shown on
Fig.1. The phase diagram contains four regions corresponding to
different phases and separated by transition lines. Each phase is
characterized by its own type of the LRO: the Neel order along $Z$
axis in the region (1); the ferromagnetic order along $Z$ axis in
the region (2); the Neel order along $Y$ axis in the region (3);
and in the region (4) there is no LRO except magnetization along
the field direction $X$ (which, certainly, exists in all above
regions). Hereafter under LRO we mean the corresponding to given
region type of LRO.

\begin{figure}[]
\unitlength=1cm
\begin{picture}(8,5.5)
\centerline{\psfig{file=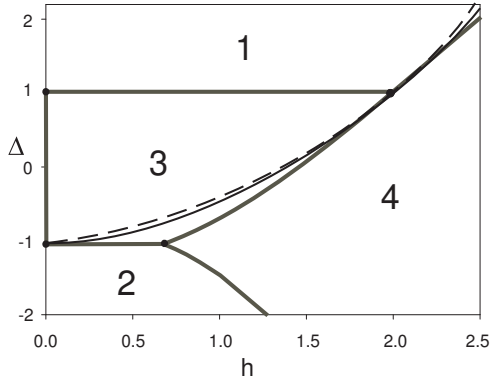,width=8cm}}
\end{picture}
\caption{Phase diagram of the model (\ref{H}). The thick solid
lines denote the critical lines, thin solid line is the
`classical' line, and dashed line denotes the line $h_1(\Delta )$
(see below).} \label{phase}
\end{figure}

In this paper we investigate the behavior of the gap and the LRO
near the transition (critical) lines. The first section is devoted
to the classical line, where we review the exact ground state and
construct three exact excitations. In the section 2 we study the
transition line $h_{c}(\Delta )$ with use of the mean-field
approach and exact diagonalization of finite systems. In the
section 3 we find the critical exponents in the vicinity of the
line $h=0$. The properties of the model near the critical lines
$\Delta =\pm 1$ and in particular in the vicinity of the points
($\Delta =\pm 1$, $h=0$) are studied in sections 4 and 5.

\section{The classical line}

First we consider the classical line $h_{\rm cl}=\sqrt{2(1+\Delta
)}$ ($\Delta >-1$), because further we shall often refer to this
line. This line is remarkable in a sense that the ground state on
it is identical to the classical one and quantum fluctuation are
missing. It was shown in \cite {classical} that the ground state
of (\ref{H}) on this line is two-fold degenerated and the ground
state wave functions with momentum $k=0$ and $k=\pi$ are
\[
\Psi _{1,2}=\frac{1}{\sqrt{2}}(\Phi _{1}\pm \Phi _{2})
\]
and $\Phi _{1(2)}$ are direct product of single-site functions:
\begin{eqnarray*}
\left| \Phi _{1}\right\rangle &=&\left| \alpha
_{1}\bar{\alpha}_{2}\alpha
_{3}\bar{\alpha}_{4}\ldots \right\rangle \\
\left| \Phi _{2}\right\rangle &=&\left| \bar{\alpha}_{1}\alpha _{2}\bar{%
\alpha}_{3}\alpha _{4}\ldots \right\rangle
\end{eqnarray*}
where $\left| \alpha _{i}\right\rangle $ is the state of ${\rm
i}$-th spin lying in the $XY$ plane for $|\Delta |<1$ (or in $XZ$
plane for $\Delta >1$) and forming an angle $\varphi $ with $X$
axis. These states can be written as:
\begin{eqnarray*}
\left| \alpha _{i}\right\rangle &=&(e^{{\rm i}\varphi
}S_{i}^{+}-1)\left|
\downarrow \right\rangle ,\quad |\Delta |<1 \\
\left| \alpha _{i}\right\rangle &=&(e^{\varphi }S_{i}^{+}-1)\left|
\downarrow \right\rangle ,\quad \Delta >1
\end{eqnarray*}
with $\cos \varphi =h_{{\rm cl}}/2$ for $|\Delta |<1$ and $\cosh \varphi =h_{%
{\rm cl}}/2$\ for $\Delta >1$.

The state $\left| \bar{\alpha}_{i}\right\rangle $ is obtained by
rotation of ${\rm i}$-th spin by $\pi $ about the axis of the
magnetic field $X$.
\[
\left| \bar{\alpha}_{i}\right\rangle =e^{{\rm i}\pi
S_{i}^{x}}\left| \alpha _{i}\right\rangle
\]

The ground state has a two-sublattice structure and is
characterized by the presence of the LRO in the $Y$ ($\left|
\Delta \right| <1)$ or in the $Z$ $(\Delta >1)$ directions. In
particular, for $|\Delta |<1$ the staggered magnetization
$\left\langle S_{n}^{y}\right\rangle $ is
\[
\left\langle S_{n}^{y}\right\rangle =\frac{(-1)^{n}}{2}\sqrt{1-\frac{h_{{\rm %
cl}}^{2}}{4}}
\]

However, the excited states of (\ref{H}) on the classical line are
nontrivial in general. Nevertheless, some of them can be found
exactly. For this aim it is convenient to introduce the operator
overturning the ${\rm i}$-th spin:
\begin{eqnarray*}
R_{i} &=&e^{{\rm i}\pi S_{i}^{z}},\quad |\Delta |<1 \\
R_{i} &=&e^{{\rm i}\pi S_{i}^{y}},\quad \Delta >1
\end{eqnarray*}
so that the states of the `overturned' ${\rm i}$-th spin $\left|
\beta
_{i}\right\rangle =R_{i}\left| \alpha _{i}\right\rangle $, $\left| \bar{\beta%
}_{i}\right\rangle =R_{i}\left| \bar{\alpha}_{i}\right\rangle $
are
orthogonal to $\left| \alpha _{i}\right\rangle $, $\left| \bar{\alpha}%
_{i}\right\rangle $:
\[
\left\langle \alpha _{i}\right| \left. \beta _{i}\right\rangle
=\left\langle \bar{\alpha}_{i}\right| \left.
\bar{\beta}_{i}\right\rangle =0
\]

Then, the exact excited state are written as:
\begin{eqnarray*}
\left| \psi _{1(2)}^{1}\right\rangle &=&\sum_{m}R_{m}\left| \Phi
_{1(2)}\right\rangle \\
\left| \psi _{1(2)}^{2}\right\rangle
&=&\sum_{n}(-1)^{n}R_{n}R_{n+1}\left|
\Phi _{1(2)}\right\rangle \\
\left| \psi _{1(2)}^{3}\right\rangle
&=&\sum_{n,m}(-1)^{n}R_{n}R_{n+1}R_{m}\left| \Phi
_{1(2)}\right\rangle
\end{eqnarray*}

So, each of three exact excitations is also two-fold degenerated.
Actually, this degeneracy is a consequence of the $Z_{2}$ symmetry
describing the rotation of all spins by $\pi $ about the axis of
the magnetic field $X$.

To show that these states are really exact ones it is convenient
to rotate the coordinate system, so that one of the ground states,
for example $\Phi _{1}$, will be all spins pointing down. For the
case $|\Delta |<1$ this transformation is the rotation of the
spins on even (odd) sites by an angle $\varphi $ $(-\varphi )$
around the $Z$ axis followed by the rotation by $\pi /2$ about the
$Y$ axis:
\begin{eqnarray}
S_{n}^{x} &=&\sigma _{n}^{z}\cos \varphi +(-1)^{n}\sigma
_{n}^{y}\sin \varphi
\nonumber \\
S_{n}^{y} &=&(-1)^{n}\sigma _{n}^{z}\sin \varphi -\sigma
_{n}^{y}\cos \varphi
\nonumber \\
S_{n}^{z} &=&-\sigma _{n}^{x}  \label{rotation1}
\end{eqnarray}

For $\Delta >1$ the transformation of the spin operators is
defined by the relation
\begin{eqnarray}
S_{n}^{x} &=&\sigma _{n}^{z}\cos \varphi +(-1)^{n}\sigma
_{n}^{x}\sin \varphi
\nonumber \\
S_{n}^{y} &=&\sigma _{n}^{y}  \nonumber \\
S_{n}^{z} &=&-(-1)^{n}\sigma _{n}^{z}\sin \varphi +\sigma
_{n}^{x}\cos \varphi  \label{rotation2}
\end{eqnarray}

Then the Hamiltonian (\ref{H}) on the classical line becomes
\begin{eqnarray}
H_{1} &=&\Delta \sum {\bf \sigma }_{n}\cdot {\bf \sigma
}_{n+1}+(1+\Delta
)\sum \sigma _{n}^{z}  \nonumber \\
&&+h_{\rm cl}\sqrt{1-\frac{h_{\rm cl}^{2}}{4}}\sum (-1)^{n}\sigma
_{n}^{y}(\sigma _{n+1}^{z}+\sigma _{n-1}^{z}+1)  \label{Hclass1}
\end{eqnarray}
for $\Delta <1$ and
\begin{eqnarray}
H_{2} &=&\sum {\bf \sigma }_{n}\cdot {\bf \sigma }_{n+1}-(\Delta
-1)\sum
\sigma _{n}^{z}\sigma _{n+1}^{z}+2\sum \sigma _{n}^{z}  \nonumber \\
&&+\sqrt{h_{\rm cl}^{2}-4}\sum (-1)^{n}\sigma _{n}^{x}(\sigma
_{n+1}^{z}+\sigma _{n-1}^{z}+1)  \label{Hclass2}
\end{eqnarray}
for $\Delta >1$.

The ground state of both Hamiltonians as well as of (\ref{H}) is
two-fold degenerated. One of the ground state is obviously all
spins ${\bf\sigma}_n$ pointing down $\Phi_1=\left| 0\right\rangle
\equiv \left| \downarrow \downarrow \downarrow \ldots
\right\rangle $. The energy of this state is:
\begin{equation}
E_{0}=-\frac{N}{2}-N\frac{\Delta }{4}  \label{Eclass}
\end{equation}

The second ground state $\Phi_2$ in this representation has a more
complicated form:
\[
\widetilde{\Phi }_{2}=\prod_{n}(\cos \varphi +(-1)^{n}\sigma
_{n}^{+}\sin \varphi )\left| 0\right\rangle
\]

Now, it is easy to see that the following three excited states are
exact ones:
\begin{eqnarray}
\left| \psi _{1}^{(1)}\right\rangle &=&\sum \sigma _{n}^{+}\left|
0\right\rangle ,\qquad E_{1}-E_{0}=1+\Delta  \label{psiclass} \\
\left| \psi _{1}^{(2)}\right\rangle &=&\sum (-1)^{n}\sigma
_{n}^{+}\sigma
_{n+1}^{+}\left| 0\right\rangle ,\quad E_{2}-E_{0}=2+\Delta  \nonumber \\
\left| \psi _{1}^{(3)}\right\rangle &=&\sum_{n,m}(-1)^{n}\sigma
_{n}^{+}\sigma _{n+1}^{+}\sigma _{m}^{+}\left| 0\right\rangle ,\
E_{3}-E_{0}=3+2\Delta  \nonumber
\end{eqnarray}

One can check that last terms in (\ref{Hclass1}) and
(\ref{Hclass2}) annihilate these three functions and, hence, they
are exact excited states of (\ref{H}) for any even $N$. Similarly
to the ground state the excited states (\ref{psiclass}) are
degenerated with the states $\left| \psi _{2}^{k}\right\rangle $.
These states $\left| \psi _{2}^{k}\right\rangle $ can be
represented in the same form (\ref{psiclass}), but in the
coordinate system, where the function $\Phi _{2}$ is all spins
pointing down.

The states $\left| \psi _{1(2)}^{1}\right\rangle $ are especially
interesting because they define the gap of the model (\ref{H}) on
the classical line at small value of $h_{{\rm cl}}$. Our numerical
calculations of finite systems show that at $h_{\rm cl}\rightarrow
0$ ($\Delta \rightarrow -1) $ the lowest branch of excitations has
a minimum at $k=0$ and corresponding excitation energy is
$(1+\Delta )$ (of course, due to the $Z_2$ symmetry there is
another branch with the minimum at $k=\pi$ and the same minimal
energy, but we consider one branch only). Excitation energy at
$k=\pi$ obtained by the extrapolation of numerical calculations at
$N\rightarrow \infty $ is $2(1+\Delta )$. When $h_{\rm cl}$
increases excitation energies at $k=0$ and $k=\pi $ are drawn
together and at some $\widetilde{h}_{\rm cl}$ they are equal to
each other. Our numerical results give $\widetilde{h}_{\rm
cl}\simeq 0.76$ ($\Delta \simeq -0.79)$. So, the gap on the
classical line is $(1+\Delta )$ for $-1<\Delta <-0.79$.

\section{The transition line $h=h_c(\Delta )$}

The existence of the transition line $h_{c}(\Delta )$ passing
through the whole phase diagram is quite natural, because at some
value of the magnetic field all types of the LRO except the LRO
along the field must vanish. The transition line connects two
obvious limits $\Delta \rightarrow \pm \infty $, when the model
(\ref{H}) reduces to the ITF model. The line passes through the
exactly solvable point ($\Delta =1$, $h=2$) and the point ($\Delta
=-1$, $h=h_{0}$) studied in \cite{Alkaraz}. We suppose that the
whole line $h_{c}(\Delta )$ is of ITF type with algebraically
decaying of correlations with corresponding critical exponents
\cite{McCoy}.

The transition line can be also observed from the numerical
calculations of finite systems. As an example, the dependencies of
excitation energies of three lowest levels on $h$ and for
$N=10-18$ are shown on Fig.2. From this figure one can see that
two lowest states cross each other $N/2$ times and the last
crossing occurs on the classical point $h_{{\rm cl}}=\sqrt{2}$.
These two states form two-fold degenerated ground state in the
thermodynamic limit. They have different momenta $k=0$ and $k=\pi
$ and different quantum number describing the $Z_{2}$ symmetry,
which remains in the system after applying the field. As for the
first excitation above the degenerated ground state, on Fig.2 we
see also numerous level crossings. These level crossings lead to
the incommensurate effects which manifest itself in the
oscillatory behavior of the spin correlation functions. The
correlation functions at $n\gg 1$ have a form:
\begin{equation}
\langle S_1^\alpha S_n^\alpha \rangle - \langle S^\alpha \rangle^2
= f(n)e^{-\kappa n} \label{incommensurate}
\end{equation}
where $\langle S^\alpha \rangle$ ($\alpha=x,y,z$) is the
corresponding magnetization (the LRO), and $f(n)$ is the
oscillatory function of $n$ with the oscillating period depending
on $h$ and $\Delta$. All crossings disappear at $h>h_{\rm
cl}(\Delta )$ and in this region of the phase diagram the
correlation functions do not contain oscillatory terms.

The energy of the first excitation near $h_{\rm cl}$ goes down
rapidly, and after extrapolation we found that for the case
$\Delta =0$ at the magnetic field $h_{c}\simeq 1.456(6)>h_{\rm
cl}$ the gap vanishes. Inside the region $h_{\rm cl}<h<h_{c}$ the
ground state remains two-fold degenerated, though there is no
level crossings. At $h>h_{c}$ the mass gap appears again and for a
large field the gap is proportional to $h$.

\begin{figure}[]
\unitlength=1cm
\begin{picture}(8,6.5)
\centerline{\psfig{file=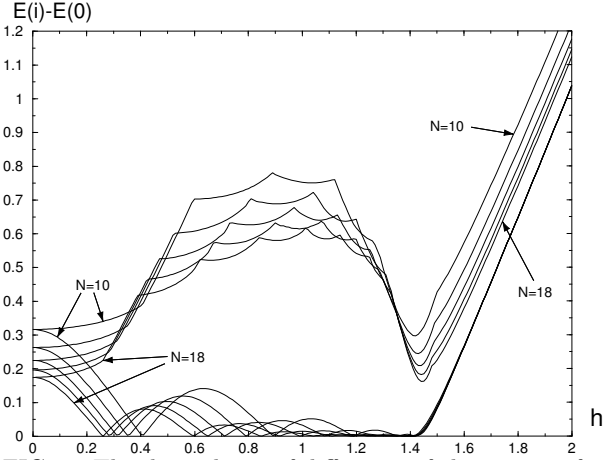,width=8cm}}
\end{picture}
\caption{The dependence of difference of the energy of two lowest
levels and the ground state energy on magnetic field $h$ for
finite chains with $N=10,\ldots 18$.} \label{abdollah}
\end{figure}

In order to determine the transition line $h_{c}(\Delta )$ and to
study the model in the vicinity of $h_{c}(\Delta )$, we use the
Fermi-representation of (\ref{H}). This representation gives the
exact solution in the limits $\Delta \rightarrow \pm \infty $ and,
besides, it yields the exact ground state on the classical line.

At first it is convenient to perform in (\ref{H}) a rotation of
the spins around the $Y$ axis by $\pi /2$, so that the magnetic
field is along the $Z$ axis:
\begin{equation}
H=\sum (\Delta
S_{n}^{x}S_{n+1}^{x}+S_{n}^{y}S_{n+1}^{y}+S_{n}^{z}S_{n+1}^{z})+h\sum
S_{n}^{z}  \label{H1}
\end{equation}

After Jordan-Wigner transformation to Fermi operators $a_{n}^{+}$,
$a_{n}$
\begin{eqnarray}
S_{n}^{+} &=&e^{i\pi \sum a_{j}^{+}a_{j}}a_{n}  \nonumber \\
S_{n}^{z} &=&a_{n}^{+}a_{n}-\frac{1}{2}  \label{JW}
\end{eqnarray}
the Hamiltonian (\ref{H1}) takes the form
\begin{eqnarray}
H_{{\rm f}} &=&-\frac{hN}{2}+\frac{N}{4}+\sum (h-1-\frac{1+\Delta
}{2}\cos
k)a_{k}^{+}a_{k}  \nonumber \\
&+&\frac{1-\Delta }{4}\sum \sin k(a_{k}^{+}a_{-k}^{+}+a_{-k}a_{k})
\nonumber
\\
&+&\sum a_{n}^{+}a_{n}a_{n+1}^{+}a_{n+1}  \label{Hfermi}
\end{eqnarray}

Treating the Hamiltonian $H_{\rm f}$ in the mean-field
approximation we find the ground state energy $E_{0}$ and the
one-particle excitation spectrum $\varepsilon (k)$:
\begin{eqnarray}
E_{0}/N &=&(h-1)\left( \gamma _{1}-\frac{1}{2}\right) +\frac{1}{4}-\left( 1-%
\frac{g}{2}\right) \gamma _{2}  \nonumber \\
&&+\frac{g}{2}\gamma _{3}+\gamma _{1}^{2}-\gamma _{2}^{2}+\gamma
_{3}^{2}
\label{Efermi} \\
\varepsilon (k) &=&\sqrt{a^{2}(k)+b^{2}(k)}  \label{spectrum}
\end{eqnarray}
where $g=1-\Delta $ and
\begin{eqnarray}
a(k) &=&(h-1)-\left( 1-\frac{g}{2}\right) \cos k+2\gamma
_{1}-2\gamma
_{2}\cos k  \nonumber \\
b(k) &=&\left( \frac{g}{2}+2\gamma _{3}\right) \sin k  \label{ab}
\end{eqnarray}

Quantities $\gamma _{1}$, $\gamma _{2}$ and $\gamma _{3}$ are the
ground state averages, which are determined by the self-consistent
equations:
\begin{eqnarray}
\gamma _{1} &=&\langle a_{n}^{+}a_{n}\rangle =\sum_{k>0}\left( 1-\frac{a(k)}{%
\varepsilon (k)}\right)  \nonumber \\
\gamma _{2} &=&\langle a_{n}^{+}a_{n+1}\rangle =-\sum_{k>0}\frac{a(k)}{%
\varepsilon (k)}\cos k  \nonumber \\
\gamma _{3} &=&\langle a_{n}^{+}a_{n+1}^{+}\rangle =-\sum_{k>0}\frac{b(k)}{%
2\varepsilon (k)}\sin k  \label{gamma}
\end{eqnarray}

Magnetization $S=\left\langle S_{n}^{x}\right\rangle $ of the model (\ref{H}%
) is equal to
\begin{equation}
S=\frac{1}{2}-\gamma _{1}  \label{Sg}
\end{equation}

The numerical solution of Eqs.(\ref{gamma}) shows that the
function $\varepsilon (k)$ has a minimum at $k_{\rm min}$, which
is changed from $\pi /2$ at $h=0$ to zero at $h=h_{1}(\Delta )$
and $k_{{\rm min}}=0$ for $h>h_{1}(\Delta )$. The gap in the
spectrum $\varepsilon (k)$ vanishes at $h_{c}(\Delta )$
($h_{c}>h_{1}$) and for $h>h_{1}$ is $m=|h-h_{c}|$. The
dependencies of $h_{1}(\Delta )$ and $h_{c}(\Delta )$ are shown on
Fig.1. We note that the Hamiltonian $H_{{\rm f}}$ differs from the
domain-wall fermionic Hamiltonian which is mapped of (\ref{H}) in
\cite{J>0}. The transition line obtained in \cite{J>0} is a linear
function of $\Delta $ in contrast to $h_{c}(\Delta )$ on Fig.1.

It is interesting to note that the mean field approximation gives
the exact ground state on the classical line $h_{\rm
cl}=\sqrt{2(1+\Delta )}$. On this line the solution of
Eqs.(\ref{gamma}) has very simple form:
\begin{eqnarray}
\gamma _{1}=\frac{1}{2}-\frac{h_{{\rm cl}}}{4}, &&\quad \quad
\gamma _{2}=-\gamma _{3}=\frac{4-h_{{\rm cl}}^{2}}{16},\qquad
\left| \Delta \right|
<1  \nonumber \\
\gamma _{1}=\frac{1}{2}-\frac{1}{h_{{\rm cl}}}, &&\quad \quad
\gamma _{2}=\gamma _{3}=\frac{h_{{\rm cl}}^{2}-4}{4h_{{\rm
cl}}^{2}},\qquad \Delta
>1  \label{gammacl}
\end{eqnarray}
and the energy $E_{0}/N=-\frac{1}{2}-\frac{\Delta }{4}$.

The gap on the classical line in the mean-field approximation is

\begin{eqnarray}
m &=&\frac{1}{4}(2-h_{{\rm cl}})^{2},\qquad \left| \Delta \right|
<1
\nonumber \\
m &=&\frac{h_{{\rm cl}}^{2}-2}{2h_{{\rm cl}}^{2}}(h_{{\rm
cl}}-2)^{2},\qquad \Delta >1  \label{mclass}
\end{eqnarray}

We compared (\ref{mclass}) with the results of the extrapolation
of finite systems on the classical line. The coincidence is fairly
good for $\Delta>0.5 $. Eq.(\ref{mclass}) gives rather
satisfactory estimation for the gap up to $\Delta \simeq -0.5$.
For example, at $\Delta =0$ ($h_{{\rm cl}}=\sqrt{2}$) from
Eq.(\ref{mclass}) $m=0.086$, while extrapolated gap is $m\simeq
0.076(4)$.

The smaller the fermion density the better works the mean-field
approximation. It becomes worse when the magnetization
$S\rightarrow 0$.\ This is the reason of incorrect behavior of the
gap at $h_{{\rm cl}}\rightarrow 0$ ($\Delta \rightarrow -1$).
According to (\ref{mclass}) $m=1$, while it vanishes in this limit
as $m=1+\Delta $ (\ref{psiclass}).

The Hamiltonian $H_{\rm f}$ in the mean-field approximation has a
form similar to the well-known bilinear Fermi-Hamiltonian
describing the anisotropic $XY$ model or the ITF model. Using
results of \cite{McCoy} the following facts related to the
considered model can be established:

1. There is a staggered magnetization $\left\langle
S_{n}^{y}\right\rangle $ along the $Y$ axis for $|\Delta |<1$ or
$\left\langle S_{n}^{z}\right\rangle $ along the $Z$ axis for
$|\Delta |>1 $ and they vanish at $h\rightarrow h_{c}$ as
$(h_{c}-h)^{1/8}$.

2. The magnetization $S$ has a logarithmic singularity at
$h\rightarrow h_{c}$.

3. The spin correlation function decay exponentially (excluding
the transition line) at $n\rightarrow\infty$:
\begin{equation}
G^{\alpha}(n) = \langle S_1^\alpha S_n^\alpha \rangle - \langle
S^\alpha \rangle^2 = f(n)e^{-\kappa n} \label{incommensurate1}
\end{equation}
The function $f(n)$ has an oscillatory behavior at $0<h<h_{\rm
cl}$, while at $h>h_{\rm cl}$ it is monotonic. At $h=h_{\rm cl}$,
$f(n)=0$ and $f(n)\sim\frac{\cos\omega n}{n^2}$
($\omega=\sqrt{2\frac{h_{\rm cl}-h}{h_{\rm cl}}}$) at $(h_{\rm
cl}-h)\ll 1$. Thus the classical line determines the boundary on
the phase diagram where the spin correlation functions show the
incommensurate behavior.

On the transition line $h=h_c(\Delta)$ the spin correlation
functions have power-law decay:
\[
G^x(n)\sim 1/n^2, \qquad G^y(n)\sim 1/n^{1/4}, \qquad G^z(n)\sim
1/n^{9/4}
\]
for $|\Delta|<1$ and
\[
G^x(n)\sim 1/n^2, \qquad G^y(n)\sim 1/n^{9/4}, \qquad G^z(n)\sim
1/n^{1/4},  \qquad |\Delta|>1   \nonumber
\]
for $|\Delta|>1$.

These results show that the transition at $h=h_{c}(\Delta )$
belongs to the universality class of the ITF model.

In the vicinity of the point $h=2$, $\Delta =1$ the fermion
density is small ($S\simeq \frac 12$) and the mean-field
approximation of the four fermion term gives the accuracy, at
least, up to $g^{3}$ or $(2-h)^{4}$. For this case we give
corresponding expressions (at $g\ll 1$):
\begin{eqnarray}
h_{c} &=&2-\frac{g}{2}-\frac{g^{2}}{32},  \nonumber \\
h_{1} &=&h_{c}-\frac{g^{2}}{16},  \nonumber \\
m &=&|h-h_{c}|,\qquad h>h_{1}  \nonumber \\
m &=&\frac{g}{2\sqrt{2}}\sqrt{h_{c}-h-\frac{g^{2}}{32}},\qquad
h<h_{1} \label{mfermi}
\end{eqnarray}

The magnetization $S$ is
\begin{eqnarray}
S &=&\frac{1}{2}-\frac{\sqrt{2}}{\pi }\sqrt{h_{c}-h}-\frac{g}{8\pi
},\qquad
g\ll \sqrt{h_{c}-h}  \nonumber \\
S &=&\frac{1}{2}-\frac{g}{4\pi }-\frac{2(h_{c}-h)}{\pi g}\ln \left( \frac{%
g^{2}}{h_{c}-h}\right) ,\qquad g\ll \sqrt{h_{c}-h}
\label{S_fermi}
\end{eqnarray}

The susceptibility $\chi (h)=\frac{dS}{dh}$ is
\begin{eqnarray}
\chi (h) &=&\frac{2}{\pi g}\ln \left( \frac{g^{2}}{h_{c}-h}\right)
,\qquad
g\gg \sqrt{h_{c}-h}  \nonumber \\
\chi (h) &=&\frac{1}{\sqrt{2}\pi }\frac{1}{\sqrt{h_{c}-h}},\qquad g\ll \sqrt{%
h_{c}-h}  \label{susfermi}
\end{eqnarray}

As follows from (\ref{susfermi}) there is a crossover from square
root to logarithmic divergence of $\chi $ when the parameter
$\frac{g^{2}}{h_{c}-h}$ is changed from $0$ to $\infty $.

\section{The line $h=0$, $|\Delta |<1$}

\subsection{Scaling estimations}

The $XXZ$ model is integrable and its low-energy properties are
described by a free massless boson field theory with the
Hamiltonian
\begin{equation}
H_{0}=\frac{v}{2}\int {\rm d}x\left[ \Pi ^{2}+(\partial _{x}\Phi
)^{2}\right] \label{Hbose}
\end{equation}
where $\Pi (x)$ is the momentum conjugate to the boson field $\Phi
(x)$, which can be separated into left and right moving terms
$\Phi =\Phi _{L}+\Phi _{R}$. The dual field $\tilde{\Phi}$ is
defined as a difference $\tilde{\Phi}=\Phi _{L}-\Phi _{R}$. The
spin-density operators are represented as
\begin{eqnarray}
S_{n}^{z} &\simeq &\frac{1}{2\pi R}\partial _{x}\Phi +{\rm const.}%
(-1)^{n}\cos \frac{\Phi }{R}  \nonumber \\
S_{n}^{x} &\simeq &\cos \left( 2\pi R\tilde{\Phi}\right) \left[ C(-1)^{n}+%
{\rm const.}\cos \frac{\Phi }{R}\right]  \label{Sboson}
\end{eqnarray}
with constant $C$ found in \cite{Lukyanov}. The compactification
radius $R$ is known from the exact solution
\[
2\pi R^{2}=\theta =1-\frac{\arccos (\Delta )}{\pi }
\]

The non-oscillating part of the operator $S^{x}$ in
Eq.(\ref{Sboson}) has scaling dimension $d=\theta /2+1/2\theta $
and conformal spin $S=1$. The non-zero conformal spin of the
perturbation operator $S^{x}$ can cause the incommensurability in
the system \cite{book}, which is in accord with
Eq.(\ref{incommensurate1}). As was shown in \cite{Nersesyan}, the
common formula for the mass gap
\begin{equation}
m\sim h^{\nu },\qquad \nu =\frac{1}{2-d}=\frac{2}{4-\theta
-1/\theta } \label{malpha}
\end{equation}
is not applicable in the whole region $|\Delta |<1$. Due to
non-zero conformal spin of the non-oscillating part of the
operator $S^{x}$ it is necessary to consider higher-order effects
in $h$. The analysis shows \cite {Nersesyan} that the original
perturbation with nonzero conformal spin generates another
perturbation with zero conformal spin
\begin{equation}
V=h^{2}\cos \left( 4\pi R\tilde{\Phi}\right)  \label{pertS1}
\end{equation}

This perturbation gives the critical exponent for the mass gap
\begin{equation}
m\sim h^{\gamma },\qquad \gamma =\frac{1}{1-\theta }
\label{mgamma}
\end{equation}

Comparing Eq.(\ref{malpha}) and (\ref{mgamma}) we see that the
perturbation (\ref{pertS1}) becomes more relevant in the region
$\Delta <\cos [\pi \sqrt{2}]\approx -0.266$.

It turns out that the oscillating part of the operator $S^{x}$
gives another, more relevant index for the gap at $\Delta <0$. Let
us reproduce usual conformal line of arguments for this
oscillating part. The perturbed action for the model is
\begin{equation}
S=S_{0}+h\int {\rm d}t{\rm d}x\,S^{x}(x,t)  \label{action}
\end{equation}
where $S_{0}$ is the Gaussian action of $XXZ$ model. The
time-dependent correlation functions of the $XXZ$ chain show the
power-law decay at $|\Delta |<1$ and have the asymptotic form
\cite{Luther}
\begin{equation}
\langle S^{x}(x,\tau )S^{x}(0,0)\rangle \sim
\frac{(-1)^{x}A_{1}}{\left( x^{2}+v^{2}\tau ^{2}\right)
^{\frac{\theta }{2}}}-\frac{A_{2}}{\left( x^{2}+v^{2}\tau
^{2}\right) ^{\frac{\theta }{2}+\frac{1}{2\theta }}} \label{XXh0}
\end{equation}
with $A_{1}$, $A_{2}$ are known constants \cite{Lukyanov} and
$\tau =it$ is imaginary time. Therefore, we can estimate the
large-distance contribution to the action of the oscillating part
of the operator $S^{x}(x,\tau )$ as
\begin{eqnarray*}
&&h\int {\rm d}\tau {\rm d}x\,S^{x}(x,\tau )\sim h\int {\rm d}\tau \sum_{n}%
\frac{(-1)^{n}}{(n^{2}+v^{2}\tau ^{2})^{\theta /4}} \\
&\sim &h\int {\rm d}\tau \sum_{n=2l}\frac{\theta n}{(n^{2}+v^{2}\tau ^{2})^{%
\frac{\theta }{4}}}\ \sim h\int {\rm d}\tau {\rm d}x\frac{\theta x}{%
(x^{2}+v^{2}\tau ^{2})^{\frac{\theta }{4}+1}}
\end{eqnarray*}

The relevant field $S^{x}(x,\tau )$ gives rise to a finite
correlation length $\xi $. This correlation length is such that
the contribution of the field $S^{x}(x,\tau )$ to the action is of
order of $1$. That is
\[
h\int_{0}^{\xi /v}{\rm d}\tau \int_{0}^{\xi }{\rm d}x\frac{\theta x}{%
(x^{2}+v^{2}\tau ^{2})^{\theta /4+1}}\sim \theta h\xi ^{1-\theta
/2}/v\sim 1
\]
which gives for the mass gap
\begin{equation}
m\sim v/\xi \sim h^{\mu },\qquad \mu =\frac{1}{1-\theta /2}
\label{mbeta}
\end{equation}

In fact, the oscillating factor $(-1)^{n}$ in the correlator in
some sense eliminates one singular integration over $x$, and into
common conformal formula $m\sim h^{\frac{1}{D-d}}$, where $D$ is
the dimension of space and $d$ is the scaling dimension of
perturbation operator, one should put $D=1$ instead of
conventional $D=2$.

The comparison of the expressions Eqs.(\ref{malpha}),
(\ref{mgamma}) and (\ref{mbeta}) shows that for $0<\Delta <1$ the
leading term is given by Eq.(\ref{malpha}), while for $-1<\Delta
<0$ by Eq.(\ref{mbeta}). Thus, one has:
\begin{eqnarray}
m &\sim &h^{\nu },\qquad 0<\Delta <1  \nonumber \\
m &\sim &h^{\mu },\qquad -1<\Delta <0  \label{mh0}
\end{eqnarray}

The functions $\nu (\Delta )$, $\mu (\Delta )$, $\gamma (\Delta )$
are shown on Fig.3.

\begin{figure}[t]
\unitlength=1cm
\begin{picture}(8,5.5)
\centerline{\psfig{file=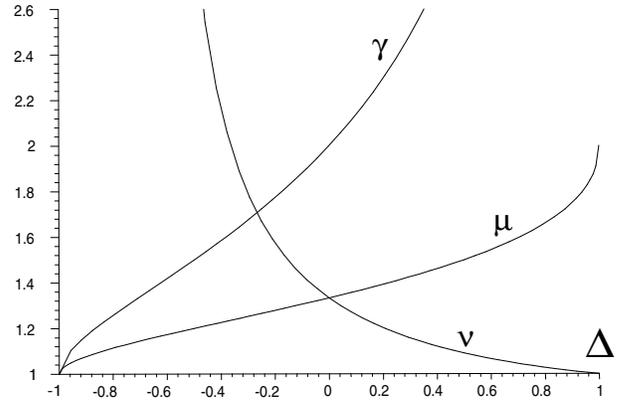,width=8cm}}
\end{picture}
\caption{The dependence of the critical exponents $\nu$, $\mu$ and
$\gamma$ on $\Delta$. The smallest exponent gives the most
relevant type of perturbation and defines the index for the mass
gap.} \label{exponents}
\end{figure}

In this respect the model (\ref{H}) is different from the $XXZ$
model in the staggered transverse field for which $m\sim
h^{2/(4-\theta )}$ for all $|\Delta |<1$ \cite{Affleck}.

The staggered magnetization (LRO) along $Y$ axis behaves as:
\begin{equation}
\left\langle S_{n}^{y}\right\rangle \sim (-1)^{n}/\xi ^{\theta
/2}\sim (-1)^{n}m^{\theta /2}  \label{LRO}
\end{equation}

Hence, the LRO has also two different critical exponents:
\begin{eqnarray}
\left\langle \left| S^{y}\right| \right\rangle &\sim &h^{\frac{\theta }{%
4-\theta -1/\theta }}\qquad 0<\Delta <1  \nonumber \\
\left\langle \left| S^{y}\right| \right\rangle &\sim &h^{\frac{\theta }{%
2-\theta }}\qquad -1<\Delta <0  \label{LROh0}
\end{eqnarray}

\subsection{Perturbation series}

The critical exponents $\nu $ and $\mu $ can be also derived from
the analysis of infrared divergences of the perturbation theory in
$h$. Obviously only even orders in $h$ gives contribution. Let us
estimate the large-distance behavior of the following operator,
defining the order of the perturbation series:
\begin{equation}
U=\frac{1}{E_{0}-H_{0}}V\frac{1}{E_{0}-H_{0}}V  \label{U}
\end{equation}
with $V=h\sum S_{i}^{x}$ and $H_{0}$ is the Hamiltonian of $XXZ$
model. The perturbation series for the ground state energy has a
form:
\begin{equation}
\delta E\sim V\frac{1}{E_{0}-H_{0}}V(1+U+U^{2}+\ldots )
\label{gsenergy}
\end{equation}

Hereafter we consider large but finite system of the length $N$.
We shall care about indices at $N$ and $h$ only, omitting all
factors.

At first we consider non-oscillating part of the correlator
(\ref{XXh0}). Taking into account only low-lying excitation of
linear spectrum of the $XXZ$ model, giving the most divergent
part, and estimating large-distance behavior of non-oscillating
part of the correlator (\ref{XXh0}), we arrive at
\begin{equation}
U\sim h^{2}\frac{\sum_{i,j}\left\langle S_{i}^{x}S_{j}^{x}\right\rangle }{%
(1/N)^{2}}\sim h^{2}N^{2}\frac{N^{2}}{N^{\theta +1/\theta }}%
=h^{2}N^{4-\theta -1/\theta }  \label{Ualpha}
\end{equation}

Now we see, that if $4-\theta -1/\theta >0$, then each next order
in perturbation theory (\ref{gsenergy})\ diverges more and more
strongly. To absorb these infrared divergences one has to
introduce the scaling parameter $y=Nh^{\nu }$ and guess that the
series $\left( 1+U+U^{2}+\ldots \right) $ in (\ref{gsenergy})
forms some function of the scaling parameter $y$. In our case $\nu
=2/(4-\theta -1/\theta )$ (see Eq.(\ref{malpha})) and $U\sim
y^{2/\nu }$.

The leading divergence of the second order of the ground state
energy can be found in a similar way as\ in (\ref{Ualpha}):
\begin{equation}
\delta E^{(2)}=V\frac{1}{E_{0}-H_{0}}V\sim h^{2}N^{3-\theta
-1/\theta } \label{dE2}
\end{equation}

Combining the formulae (\ref{Ualpha}) and (\ref{dE2}),\ we can
rewrite
\[
\delta E\sim Nh^{2\nu }f(y)
\]
with some unknown function $f(y)$ having the expansion at small
$y$:
\[
f(y)=\frac{1}{y^2}\sum_{n=1}^{\infty }c_{n}y^{2n/\nu }
\]

In the thermodynamic limit $N\rightarrow \infty $ the scaling
parameter $y=Nh^{\nu }$ also tends to infinity $y\rightarrow
\infty $. Since the energy is proportional to $N$, the function
$f(y)$ has a finite limit $f(\infty )=a$. Thus, in the
thermodynamic limit for the correction to the ground state energy
one has:
\begin{equation}
\delta E\sim aNh^{2\nu }  \label{dEalpha}
\end{equation}

For the first excitation state the divergences of the orders of
the perturbation theory have the same form as in (\ref{Ualpha})
and (\ref{dE2}). So, for the gap one finds the same scaling
parameter $y=Nh^{\nu }$ and
\[
m\sim Nh^{2\nu }g(y)
\]

In the thermodynamic limit the mass gap is of order of unity (in
terms of $N$), and, therefore, the function $g(y)\sim 1/y$ at
$y\rightarrow \infty $. Thus, finally we arrive at the
Eq.(\ref{malpha}).

Now we consider more subtle, oscillating part of the correlator
(\ref{XXh0}). For the oscillating part at large distances one can
write:
\[
\sum_{i,j}\left\langle S_{i}^{x}S_{j}^{x}\right\rangle \sim N\sum_{r}\frac{%
(-1)^{r}}{r^{\theta }}\sim N\sum_{r}\frac{1}{r^{\theta +1}}\sim N\frac{1}{%
N^{\theta }}
\]

It turns out that the oscillating part of the perturbation
operator $V$ connects the low-lying gapless states with the finite
energy states. That is, each second level in all orders of the
perturbation series has a finite gap from the ground state.
Therefore, for the operator $U$ one has
\[
U\sim h^{2}\frac{\sum_{i,j}\left\langle S_{i}^{x}S_{j}^{x}\right\rangle }{%
(1/N)}\sim h^{2}N^{2-\theta }
\]

Since $\theta $ is always less than $2$ the divergences grow with
the order of the perturbation theory. To eliminate these
divergences we introduce the scaling parameter $y=Nh^{\mu }$, with
$\mu $ defined in Eq.(\ref{mbeta}), so that $U\sim y^{2/\mu }$.

The second order for the ground state energy in this case is
\[
\delta E^{(2)}\sim h^{2}\frac{\sum_{i,j}\left\langle
S_{i}^{x}S_{j}^{x}\right\rangle }{1}\sim h^{2}N^{1-\theta }
\]
and the total correction to the ground state energy has a form
\[
\delta E\sim Nh^{2\mu }f(y)
\]
with some unknown function $f(y)$, having the finite limit
$f(\infty )=b$.

Thus, in the thermodynamic limit the ground state energy behaves
as
\[
\delta E\sim bNh^{2\mu }
\]

The mass gap is found similarly
\[
m\sim Nh^{2\mu }g(y)
\]
with the function $g(y)\sim 1/y$ at $y\rightarrow \infty $. Thus,
in the thermodynamic limit we reproduce Eq.(\ref{mbeta}).

We note that we have estimated only long wave-length divergent
part of the perturbation theory. Besides, there is a regular part
of the perturbation theory, which gives the leading term $\sim
h^{2}$. So, combining all the above facts we arrive at
\begin{equation}
\frac{\delta E}{N}=-\frac{\chi }{2}h^{2}+ah^{2\nu }+bh^{2\mu }
\label{dE}
\end{equation}

As it can be seen from Eq.(\ref{dE}), $\delta E$ consists of a
regular term ~$h^{2}$ and two singular terms. Since $\nu >1$ and
$\mu >1$, the susceptibility $\chi $ is finite for any $\Delta $
in contrast to the model with staggered transverse field
\cite{Affleck}, where the singular term is~$h^{\eta }$ with $\eta
=4/(4-\theta )<2$.

From Eq.(\ref{malpha}) and (\ref{mbeta}) it follows that $\nu
\rightarrow 1$ at $\Delta \rightarrow 1$ and $\mu \rightarrow 1$
at $\Delta \rightarrow -1$. Hence, in both limits one of the
singular terms becomes proportional to $h^{2}$, and, therefore,
gives contribution to the susceptibility. It implies that in the
symmetric points $\Delta =\pm 1$ the susceptibility has a jump.

\section{The line $\Delta =1$}

In the vicinity of the line $\Delta =1$ it is convenient to
rewrite the Hamiltonian (\ref{H}) in the form
\begin{eqnarray}
H &=&H_{0}+V  \nonumber \\
H_{0} &=&\sum {\bf S}_{n}\cdot {\bf S}_{n+1}+h\sum S_{n}^{x}  \nonumber \\
V &=&-g\sum S_{n}^{z}S_{n+1}^{z}  \label{Hd1}
\end{eqnarray}
where the parameter $g=1-\Delta \ll 1$ is small. On the isotropic
line $\Delta =1$ the model (\ref{H}) is exactly solvable by Bethe
ansatz. The properties of the system remain critical up to the
transition point $h_{c}=2$, where the ground state becomes
ferromagnetic. Therefore, for $h<2$ and small perturbation $V$ we
can use conformal estimations.

The asymptotic of the correlation function on this line is
\begin{equation}
\langle S_{i}^{z}S_{i+n}^{z}\rangle \sim \frac{(-1)^{n}}{n^{\alpha
(h)}}, \label{ZZd1}
\end{equation}
where $\alpha (h)$ is known function obtained from Bethe ansatz
\cite{BIK} and having the following limits:
\begin{eqnarray}
\alpha (h) &\sim &1-\frac{1}{2\ln \left( 1/h\right) },\qquad
h\rightarrow 0
\nonumber \\
\alpha (h) &\sim &\frac{1}{2},\qquad h\rightarrow 2
\label{alphah}
\end{eqnarray}

So, the scaling dimension of operator $S^{z}$ is $d_{z}=\alpha
(h)/2$, and the scaling dimension of operator
$S_{i}^{z}S_{i+1}^{z}$ is four times greater
$d_{zz}=4d_{z}=2\alpha (h)$. Since $\alpha (h)<1$, then the
perturbation $V$ is relevant and leads to the mass gap and the
staggered magnetization
\begin{eqnarray}
m &\sim &\left| g\right| ^{1/(2-d_{zz})}=\left| g\right|
^{1/(2-2\alpha )}
\nonumber \\
\left\langle \left| S^{y}\right| \right\rangle &\sim &\left|
g\right|
^{\alpha /(4-4\alpha )},\qquad \Delta <1  \nonumber \\
\left\langle \left| S^{z}\right| \right\rangle &\sim &\left|
g\right| ^{\alpha /(4-4\alpha )},\qquad \Delta >1  \label{md1}
\end{eqnarray}

From the general expressions for the mass gap (\ref{md1}), in the
limit $h\rightarrow 2$ we obtain that $m\sim g$, which is in
accord with the result of the mean field approximation
(\ref{mfermi}).

The LRO in the vicinity of the point $\Delta =1$, $h=2$ vanishes
on both lines: at $\Delta =1$ from (\ref{md1}) as $g^{1/4}$; and
at $h=h_{c}$ as $\left| h_{c}-h\right| ^{1/8}$. Besides, one has
the exact expression for LRO on the classical line
\begin{equation}
\left\langle \left| S^{y}\right| \right\rangle _{{\rm cl}}=\frac{\sqrt{g}}{2%
\sqrt{2}}  \label{LROcl}
\end{equation}
Combining all these facts we arrive at the following formula:
\begin{equation}
\ \ \left\langle \left| S^{y}\right| \right\rangle
=2^{-7/8}g^{1/4}\left| h_{c}-h\right| ^{1/8}  \label{LROd1h2}
\end{equation}

The behavior of the system near the point $\Delta =1$, $h=0$ is
more subtle. As it follows from Eq.(\ref{mh0}), for very small $h$
the mass gap is $m\sim h$, while on the other hand from
Eq.(\ref{md1}) one obtains another scaling $m\sim g^{\ln (1/h)}$.
So, there are two regions near this point with different behavior
of mass gap. The boundary between these two regions can be found
from the following consideration. Let us rewrite the perturbation
in the Hamiltonian (\ref{Hd1}) in the form:
\begin{eqnarray*}
V &=&V_{1}+V_{2} \\
V_{1} &=&-\frac{g}{2}\sum \left(
S_{n}^{y}S_{n+1}^{y}+S_{n}^{z}S_{n+1}^{z}\right) \\
V_{2} &=&\frac{g}{2}\sum \left(
S_{n}^{y}S_{n+1}^{y}-S_{n}^{z}S_{n+1}^{z}\right)
\end{eqnarray*}

The part of the Hamiltonian $H_{0}+V_{1}$ corresponds to the $XXZ$
model in the longitudinal magnetic field, which is gapless when
the magnetic field $h>\exp \left( -\pi ^{2}/2\sqrt{g}\right) $.
Therefore, in the region of very small magnetic field $h<\exp
\left( -\pi ^{2}/2\sqrt{g}\right) $ the perturbation $V_{1}$ is
relevant, leading to the mass gap $m\sim h$. The two-cutoff
scaling procedure \cite{book}\cite{Nersesyan} gives rise to the
mass gap $m\sim h\exp \left( -\pi ^{2}/2\sqrt{g}\right) $ for
$h>\exp \left( -\pi ^{2}/2\sqrt{g}\right) $. And, finally, when
$g$ is much less than $h$, the scaling dimension of operator
$V_{2}$ defines the exponent for the gap (\ref{md1}). Summarizing
all above, the mass gap in the vicinity of the isotropic point
$\Delta =1$, $h=0$ is:
\begin{eqnarray}
m &\sim &h,\qquad \qquad \ln h\ll -\frac{1}{\sqrt{g}}  \nonumber \\
m &\sim &he^{-\pi ^{2}/2\sqrt{g}},\qquad \frac{1}{\sqrt{g}\ln
g}\gg \ln h\gg
-\frac{1}{\sqrt{g}}  \nonumber \\
m &\sim &g^{-\ln h},\qquad \qquad \ln h\gg \frac{1}{\sqrt{g}\ln g}
\label{md1h0}
\end{eqnarray}

\section{The line $\Delta =-1$}

In this section we consider the model (\ref{H}) in the vicinity of
the line $\Delta =-1$, where $(1+\Delta )=\delta \ll 1$ is a small
parameter. It is convenient to rotate spins on each odd site by
$\pi $ around $Z$ axis, so that the model (\ref{H}) becomes:
\begin{equation}
H=-\sum {\bf S}_{n}\cdot {\bf S}_{n+1}+\delta \sum
S_{n}^{z}S_{n+1}^{z}-h\sum (-1)^{n}S_{n}^{x}  \label{Hd-1}
\end{equation}

At $\delta =0$ and $h=0$ the ground state of \ (\ref{Hd-1}) is
ferromagnetic state (degenerated with respect to total $S^{z}$)
with zero momentum. The states which can be reached from the
ground state by means of the transition operator $\sum
(-1)^{n}S_{n}^{x}$ are the states with $q=\pi $ and finite gap
over the ground state. When $\delta \ll 1$ the transition operator
connects the low-energy states and the states with energies
$\varepsilon _{s}\simeq 2$. The second order correction to
low-energy states is
\begin{equation}
\delta E_{l}^{(2)}=h^{2}\sum_{s,n,m}\frac{\left\langle l\right|
(-1)^{n}S_{n}^{x}\left| s\right\rangle \left\langle s\right|
(-1)^{m}S_{m}^{x}\left| l\right\rangle }{E_{l}-E_{s}}
\label{dEl2a}
\end{equation}
where $\left| l\right\rangle $ is the low-energy state and $\left|
s\right\rangle $ is the state with high energy $\simeq 2$. So, for
$\delta \ll 1$ Eq.(\ref{dEl2a}) can be rewritten as
\begin{eqnarray}
\delta E_{l}^{(2)} &=&-\frac{h^{2}}{2}\sum_{nm}\left\langle
l\right|
(-1)^{n-m}S_{n}^{x}S_{m}^{x}\left| l\right\rangle  \label{dEl2b} \\
&&=-\frac{h^{2}N}{8}-h^{2}\sum_{n<m}\left\langle l\right|
(-1)^{n-m}S_{n}^{x}S_{m}^{x}\left| l\right\rangle  \nonumber
\end{eqnarray}

The spin correlation function $\left\langle l\right|
S_{n}^{x}S_{m}^{x}\left| l\right\rangle $ is slow varying function
of $\left| m-n\right| $ at $\delta \ll 1$. Therefore,
\begin{equation}
\sum_{n<m}\left\langle l\right| (-1)^{n-m}S_{n}^{x}S_{m}^{x}\left|
l\right\rangle \simeq -\frac{1}{2}\sum_{n}\left\langle l\right|
S_{n}^{x}S_{n+1}^{x}\left| l\right\rangle  \label{dEl2c}
\end{equation}

Thus, as follows from Eqs.(\ref{dEl2a}-\ref{dEl2c}), the low-lying
states of (\ref{Hd-1}) at $\left| \delta \right| \ll 1$ and $h\ll
1$ are described by the $XYZ$ Hamiltonian
\begin{equation}
H=-\frac{h^{2}N}{8}-\sum [(1-\frac{h^{2}}{2}%
)S_{n}^{x}S_{n+1}^{x}+S_{n}^{y}S_{n+1}^{y}-\Delta
S_{n}^{z}S_{n+1}^{z}] \label{HXYZ}
\end{equation}

The coincidence of the low-energy spectra of (\ref{Hd-1}) and
(\ref{HXYZ}) in the vicinity of ferromagnetic point $\Delta =-1$,
$h=0$ has been checked numerically for finite systems. The
spectrum of low-lying excitations of the $s=\frac{1}{2}$ $XYZ$
model (\ref{HXYZ}) as well as initial model (\ref{H}) near the
ferromagnetic point $\Delta =-1$, $h=0$ can be described
asymptotically exactly by the spin-wave theory, which gives
\begin{eqnarray}
m &=&h\sqrt{(1+\Delta )/2},{\rm \qquad }\Delta >-1  \label{mXYZ} \\
m &=&\sqrt{(1+\Delta )(1+\Delta +h^{2}/2)},{\rm \qquad }\Delta <-1
\nonumber
\end{eqnarray}

It can be checked that Eq.(\ref{mXYZ}) yields the exact gap of the
$XYZ$ model \cite{kosevich} at $\left| \delta \right| $, $h\ll 1$.
The validity of the spin-wave approximation is quite natural
because in the vicinity of the ferromagnetic point $\Delta =-1$,
$h=0$ the number of magnons forming the ground state is small.

We note also that the gap (\ref{mXYZ}) for $\Delta \geq -1$ is in
accord with the conformal theory result (\ref{mh0}) and gives us
the preexponential factor for the gap. On the classical line
$h_{{\rm cl}}=\sqrt{2(1+\Delta )}$ \ Eq.(\ref{mXYZ}) yields the
gap $m=1+\Delta $ which confirms that the function $\psi_1^{(1)}$
in Eq.(\ref{psiclass}) gives the exact gap.

The similar mapping of the model (\ref{H}) with arbitrary spin $s$
to $XYZ$ model can be done for $\Delta \simeq -1$, $h\ll 1$.
Taking into account that $\varepsilon _{s}=4s$ corresponding $XYZ$
Hamiltonian is
\begin{eqnarray}
H=-\sum
[(1-\frac{h^{2}}{2})S_{n}^{x}S_{n+1}^{x}+S_{n}^{y}S_{n+1}^{y}-\Delta
S_{n}^{z}S_{n+1}^{z}] \nonumber\\
- \frac{h^{2}}{4s}\sum (S_{n}^{x})^{2} \label{HXYZs}
\end{eqnarray}
where $S_{n}^{\alpha }$ are spin-$s$ operators.

The leading term of the gap of the model (\ref{H}) with arbitrary
spin $s$ in the vicinity of the point $\Delta =-1$, $h=0$ is given
exactly by the spin-wave theory:

\begin{eqnarray}
m &=&h\sqrt{(1+\Delta )/2},{\rm \qquad }\Delta >-1  \label{mXYZs} \\
m &=&2s\sqrt{(1+\Delta )(1+\Delta +h^{2}/8s^{2})},{\rm \qquad
}\Delta <-1 \nonumber
\end{eqnarray}

On the classical line $h_{{\rm cl}}=2s\sqrt{2(1+\Delta )}$
Eq.(\ref{mXYZs}) gives the correct result $m=2s\delta $.

Strictly on the line $\Delta =-1$ the model (\ref{H}) reduces to
the isotropic ferromagnet in the staggered magnetic field. This
model is non-integrable, but it was suggested \cite{Alkaraz}, that
the system is governed by a $c=1$ conformal field theory up to
some critical value $ h=h_{0} $, where the phase transition of the
Kosterlitz-Thouless type takes place.

At $h\ll 1$ where the mapping of (\ref{Hd-1}) to the $XYZ$ model
(\ref{HXYZ}) is valid, the line $\Delta =-1$ is described by the
$XXZ$ model and the correlation functions have a power-law decay:

\begin{eqnarray}
\langle S_{i}^{z}S_{i+n}^{z}\rangle &=&\langle
S_{i}^{y}S_{i+n}^{y}\rangle
\sim \frac{(-1)^{n}}{n^{1/\beta (h)}}  \nonumber \\
\langle S_{i}^{x}S_{i+n}^{x}\rangle &\sim
&\frac{(-1)^{n}}{n^{\beta (h)}} \label{corrd-1}
\end{eqnarray}

We believe that the relation between indices of $X$ and $Y$, $Z$
correlators on the line $\Delta =-1$ has the form (\ref{corrd-1})
for $0<h<h_{0}$. So, the scaling dimensions of operators
$S_{i}^{x}$ and $S_{i}^{y}$, $S_{i}^{z}$ on this line are
$d_{x}=\beta /2$ and $d_{y}=d_{z}=1/2\beta $.

On the line $\Delta =-1$ the model (\ref{H}) is gapless for
$h<h_{0}$. It means that the magnetic field term is irrelevant at
$h<h_{0}$ ($\beta (h)>4$) and becomes marginal at $h=h_{0}$, where
$d_{x}=2$ and $\beta (h_{0})=4$. So, at the point $h=h_{0}$ the
transition is of the Kosterlitz-Thouless type, and for $h>h_{0}$
the mass gap is exponentially small.

In the vicinity of the line $\Delta =-1$ the term $\delta \sum
S_{n}^{z}S_{n+1}^{z}$ in (\ref{Hd-1}) can be considered as a
perturbation and the scaling dimension of the perturbation
operator $S_{n}^{z}S_{n+1}^{z}$ is $d_{zz}=4d_{z}=2/\beta (h)$.
Since $\beta (h)\geq 4$ at $h<h_{0}$, the perturbation is relevant
and leads to the mass gap and the LRO:
\begin{eqnarray}
m &\sim &\left| \delta \right| ^{\frac{1}{2-2/\beta }}  \nonumber \\
\left\langle \left| S^{y}\right| \right\rangle &\sim &\delta ^{\frac{1}{%
4(\beta -1)}},\quad \quad \delta >0  \nonumber \\
\left\langle \left| S^{z}\right| \right\rangle &\sim &\left|
\delta \right| ^{\frac{1}{4(\beta -1)}},\quad \quad \delta <0
\label{md-1}
\end{eqnarray}

In particular, $m\sim \left| \delta \right| ^{2/3}$ and
$\left\langle \left| S^{y}\right| \right\rangle \sim \left| \delta
\right| ^{1/12}$ at $h\rightarrow h_{0}$.

The function $\beta (h)$ is generally unknown, except the case
$h\ll 1$, where the mapping to $XXZ$ model is valid, and $\beta
(h)=\left[ 1-\frac{1}{\pi }\arccos (\frac{h^{2}}{2}-1)\right]
^{-1}\simeq \pi /h$. However, since the model (\ref{H}) at $\Delta
=-1$ and $h<h_{0}$ is conformal invariant, we can use a
finite-size scaling analysis to determine the exponent $\beta (h)$
and the value of $h_{0}$. According to the standard scaling
approach \cite{finitesize} $\beta (h)=\frac{2\pi v}{A}$, where $v$
is the sound velocity and $\frac{A}{N}$ is the difference between
the two lowest energies of the system. We carried out calculations
of $\beta (h)$ for finite systems. The extrapolated function
$\beta (h)$ agrees well with the dependence $\pi /h$ at $h\ll 1$
and $\beta =4$ at $h_{0}\simeq 0.52$. This estimation is close to
our direct numerical estimations $h_{0}\simeq 0.549$. On the other
hand, the mean-field approach gives rather crude value
$h_{0}=h_{c}(-1)=0.69$.

\section{Conclusion}

In summary, we have studied the effect of the symmetry-breaking
transverse magnetic field on the $s=1/2$ $XXZ$ chain. On the
contrary to the longitudinal field the transverse field generates
the staggered magnetization in $Y$ direction and the gap in the
spectrum of the model with the easy-plain anisotropy. Using the
conformal invariance we have found the critical exponents of the
field dependence of the gap and the LRO. It was shown that the
spectrum of the model is gapped on the whole ($h$, $\Delta $)
plain except several critical lines, where the gap and the LRO
vanish. The behavior of the gap and the LRO in the vicinity of the
critical lines $\Delta =\pm 1$ is considered on the base of the
conformal field theory. We note that in the vicinity of the points
($\Delta =1$, $h=0$) and ($\Delta =1$, $h=2$) there is crossover
between different regimes of the system behavior. It is shown that
near the point ($\Delta =-1$, $h=0$) the initial model can be
mapped to the effective exactly solvable 1D $XYZ$ model and has
the spin-wave spectrum. The transition line $h_{c}(\Delta )$
between the ordered phases and the disordered one is studied in
the mean-field approximation. This study shows that this
transition is similar to that in the Ising model in the transverse
field. However, the behavior of the model on the transition line
near the Kosterlitz-Thouless point ($\Delta =-1$, $h=h_{0}$) is
not so clear. The mean-field approximation becomes worse at
$\Delta \rightarrow -1$ and more sophisticated theory is needed.

We thank Prof. P.Fulde for many useful discussions. We are
grateful to Max-Planck-Institut fur Physik Komplexer Systeme for
kind hospitality. This work is supported under RFFR Grants No
00-03-32981 and No 00-15-97334.

\end{document}